\documentclass[aip,reprint,groupedaddress]{revtex4-1}
\usepackage{graphicx}
\usepackage{dcolumn}
\usepackage{bm}
\usepackage [english]{babel}
\usepackage [autostyle, english = american]{csquotes}
\MakeOuterQuote{"}
\usepackage{hyperref}
\hypersetup{colorlinks,linkcolor={red},citecolor={red},urlcolor={red}}
\usepackage {amssymb}
\usepackage[fleqn]{amsmath,nccmath}
\usepackage{environ}

\usepackage[normalem]{ulem}	
\usepackage{cancel}			
\usepackage{xfrac}			
\usepackage{color}			
\usepackage[euler]{textgreek}	

\begin{document}


\title{Nanoscale virtual potentials using optical tweezers}
\author{Avinash Kumar}
\email{aka106@sfu.ca}
\author{John Bechhoefer}
\email{johnb@sfu.ca}
\affiliation{
Dept.~of Physics, Simon Fraser University, 8888 University Dr., Burnaby, BC, V5A 1S6, Canada}

\begin{abstract}
We combine optical tweezers with feedback to impose arbitrary potentials on a colloidal particle. The feedback trap detects a particle's position, calculates a force based on an imposed ``virtual potential,'' and shifts the trap center to generate the desired force.  We create virtual harmonic and  double-well potentials to manipulate particles.  The harmonic potentials can be chosen to be either weaker or stiffer than the underlying optical trap.  Using this flexibility, we create an isotropic trap in three dimensions.  Finally, we show that we can create a virtual double-well potential with fixed well separation and adjustable barrier height.  These are accomplished at length scales down to 11 nm, a feat that is difficult or impossible to create with standard optical-tweezer techniques such as time sharing, dual beams, or spatial light modulators.   
\end{abstract}

\maketitle

\noindent Over the last three decades, optical tweezers\cite{ashkin1986observation,jones15} have been used  to exert piconewton forces on mesoscopic particles and detect their motion for physical,\cite{henderson2001direct,liphardt2002equilibrium,wang2002experimental,berut2012experimental} chemical,\cite{yao1996optical,ajito2002single} and biological applications.\cite{ashkin1989internal,block1990bead,wang1997stretching,woodside2006direct}  In parallel with these applications of optical tweezers, feedback forces have been another way to trap particles and exert small forces.  Although the details of such \textit{feedback traps} vary, they share the common feature of operating in a cycle where one measures the position of a particle, calculates a desired trapping force, and then applies it (Fig. \ref{eqn:update rule}).  Often the goal is simply to trap an object, a task that has been done using many different types of force for the feedback, including electrokinetic,\cite{cohen05b} magnetic,\cite{gosse2002magnetic} microfluidic flow,\cite{armani2006using} and thermophoretic forces.\cite{braun2013optically,lin2018opto}  The objects trapped have ranged from colloidal particles to bacteria to proteins and even to individual dye molecules diffusing in water.\cite{fields11}  Trapping allows one to measure, with good statistics, physical properties of individual objects.\cite{wang11}   In other situations, the goal is not simply to trap but to create a more-complicated force field, for example a \textit{virtual potential} that can be a discrete approximation to a physical potential,\cite{cohen2005control} an idea that has been used to test fundamental aspects of statistical physics such as the relations between information and thermodynamics,\cite{jun2014high,gavrilov16b} or the measurement of the functional form of the Gibbs-Shannon entropy function.\cite{gavrilov17b}

Feedback has been used previously in optical tweezers, but for relatively simple goals such as increasing the stiffness of the trap relative to its normal value.  Simmons et al. \cite{simmons1996quantitative} achieved a 400-fold gain in the stiffness using two-dimensional analog feedback control provided by a pair of orthogonal acousto-optic deflectors (AODs).  Using similar setups based on digital feedback control, Ranaweera et al.\cite{ranaweera2005modelling} and Wallin et al.\cite{wallin2008stiffer} achieved 29-fold and 10-fold gains in stiffness, respectively.

We create virtual harmonic potentials and double-well potentials.  In previous studies on feedback traps,\cite{cohen2005control,jun2014high} such virtual potentials were created by applying electrokinetic forces, which are particularly well suited for applying strong forces to nanometer-scale particles.\cite{fields11}  Here, we substitute the dielectrophoretic forces due to optical tweezers.  

\begin{figure}[tbh]
	\includegraphics[width=2.5in]{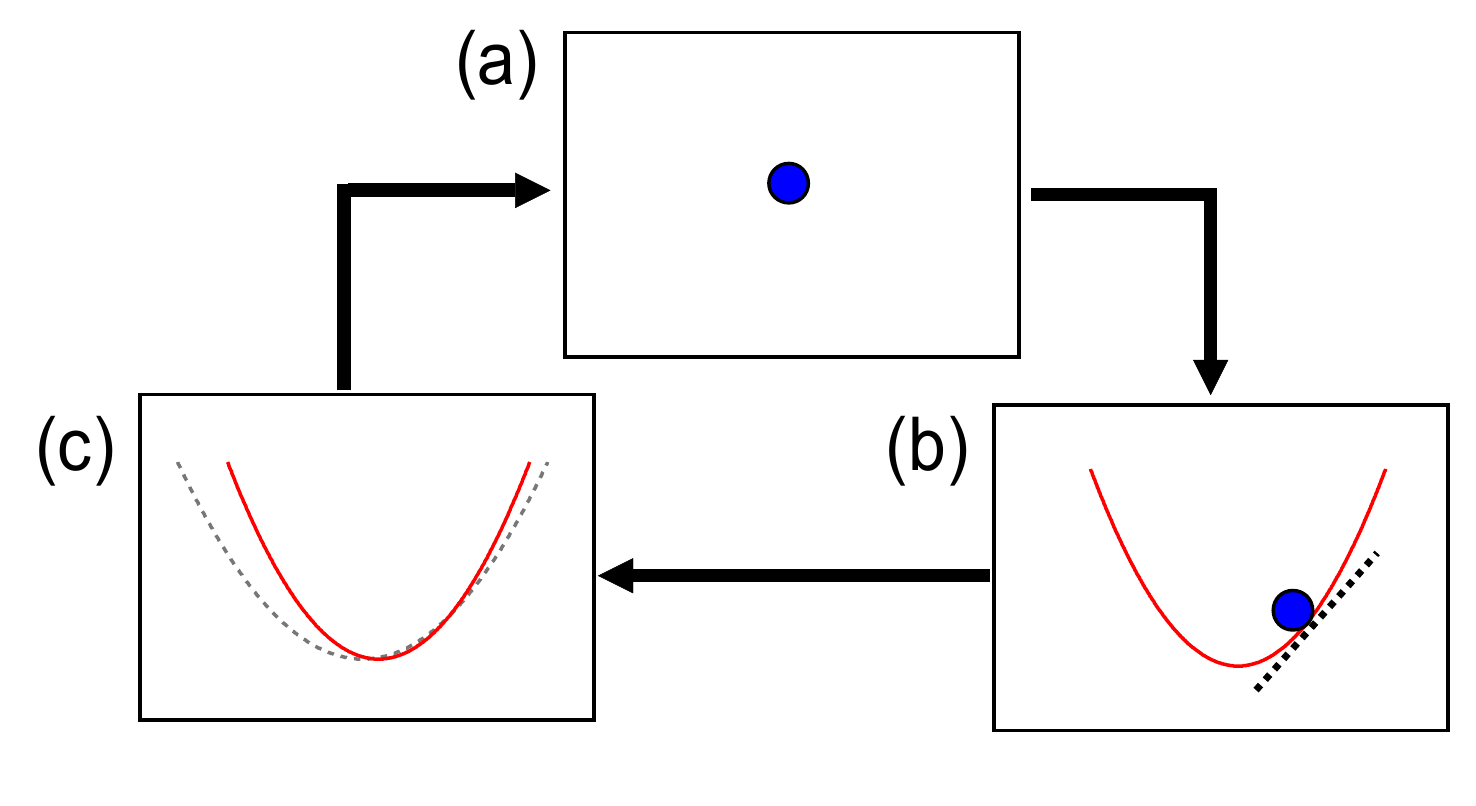}
\caption[FB_rule] { \label{fig:FB_rule} 
One cycle of the feedback trap: (a) acquire information about particle's position, ($x$,$y$); (b) calculate the force based on the chosen virtual potential; and (c) move the trap position to generate equivalent force. The harmonic potential in red in (b) is the chosen potential and the gray dashed harmonic potential in (c) is the underlying tweezer potential, which is shifted to exert a force on the particle.}
\end{figure}

By countering the Brownian fluctuations with feedback using the force from optical tweezers, we obtain a 30-fold gain in the stiffness of the trap, keeping the laser power constant in the trapping plane.  In contrast to previous work,\cite{wallin2008stiffer} we can also \textit{reduce} the trapping strength using feedback. We take advantage of this flexibility to create an isotropic trap using feedback in both X and Y directions. To further show the ability of feedback traps to form arbitrary potentials, we construct a double-well potential with well separation and well curvatures that are held fixed but with a barrier height that can be set as desired.

\begin{figure}[h]
	\includegraphics[width=3.2in]{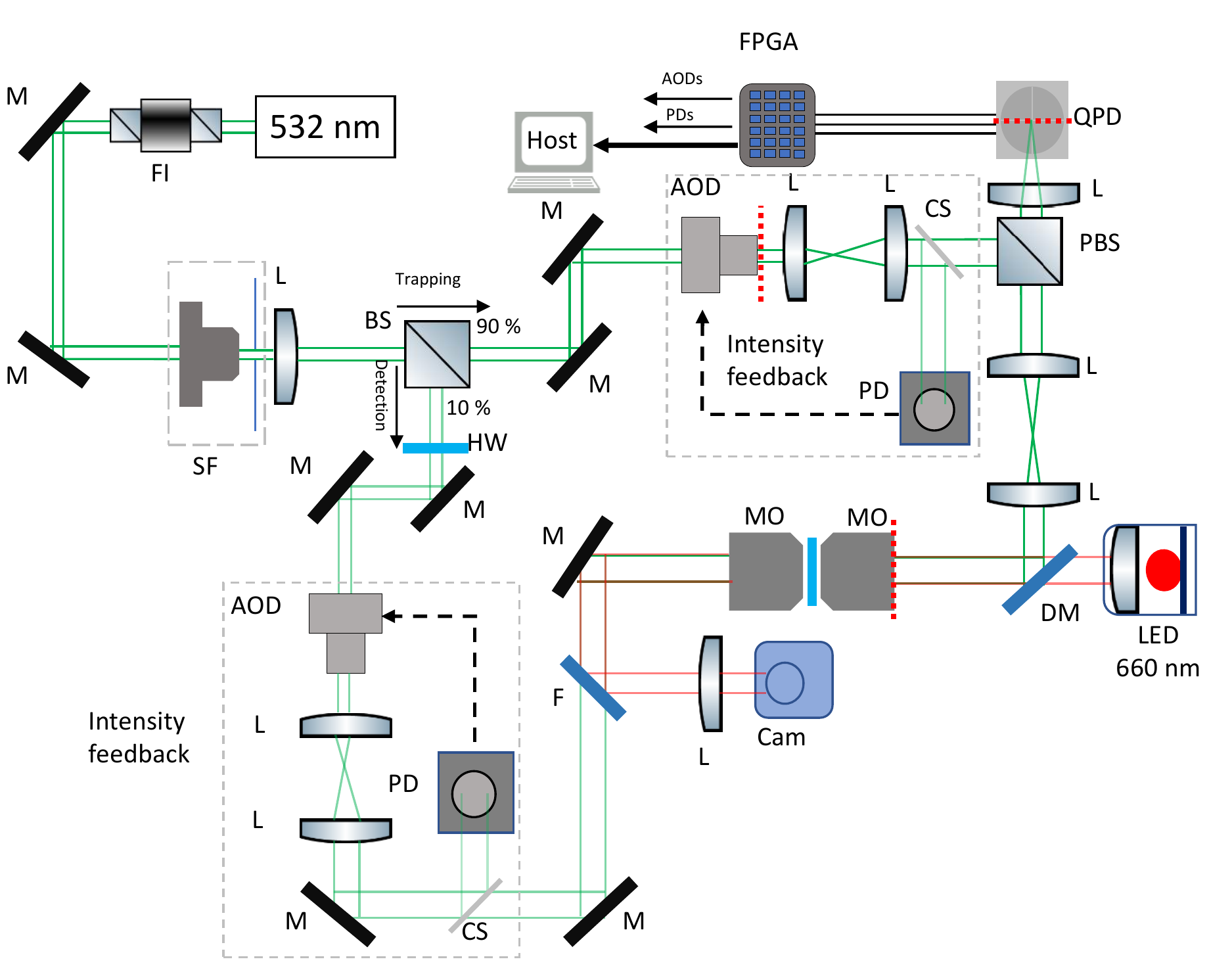}
	\caption[setup] { \label{fig:setup} 
Schematic diagram of the feedback-trap setup. FI = Faraday Isolator, M = Mirror, SF = Spatial Filter, BS = Beam Splitter (non-polarizing), AOD = Acousto-Optic Deflector, L = Lens, MO = Microscope Objective, SC = Sample Chamber, PBS = Polarizing Beam Splitter, HW = Half-Wave Plate, F = Short-Pass Filter, QPD = Quadrant Photodiode, DM = Dichroic Mirror, PD = Photodiode, CS = Cover-Slip, Cam = Camera. Planes conjugate to the back-focal plane of the trapping objective are shown in red-dashed lines.}
\end{figure}

The tweezer-based feedback trap is based on a custom-built microscope constructed on a vibration-isolation table (Melles Griot) (Fig. \ref{fig:setup}). An $s$-polarized 532 nm laser (Nd:YAG, Coherent Genesis MX STM-series, 1 Watt) is used for trapping and detection. The laser passes through a Faraday isolator (LINOS FI-530-2SV), which protects the laser cavity from back-reflections. The laser beam is separated into trapping and detection beams using a 90:10 beam splitter. The polarization of the detection beam is rotated by 90$^{\circ}$ with a half-wave plate to minimize any interference with the trapping laser. Both beams pass independently through two AODs that provide orthogonal XY deflection (DTSXY-250-532, AA Opto Electronic).  Each AOD can change the intensity in the first-order diffraction beam and steer its angle using an analog voltage controlled oscillators (DFRA10Y-B-0-60.90, AA Opto Electronic). The beams are then expanded by a factor of two to overfill the back aperture of the microscope objectives.  Relay lenses image the steering point of the AOD onto the back focal plane of the trapping objective, to translate beam rotation into linear motion in the trapping plane. 

We use two identical water-immersion high-numerical-aperture objectives (Olympus 60X, UPlanSApo, NA  = 1.20 W) for trapping and detection of a $1.49\,\mu$m spherical silica bead (Bangs Laboratories).  An XY-piezo stage (Mad City Labs, H100) is attached to the trapping objective to provide a precise movement of the trap inside the sample chamber. The trapping objective also collects the forward scattered light from the detection beam. A polarized beam splitter (PBS25-532-HP, Thorlabs) separates the detection beam and the back-scattered light arising from the trapping laser by transmitting the former and reflecting the latter.  We use a quadrant photodiode (QPD, First Sensor, QP50-6-18u-SD2) to detect the particle's fluctuations. The QPD is placed at the back focal plane of the trapping objective for back-focal-plane interferometry.\cite{gittes1998interference}  A 660 nm LED (Thorlabs, M660L4) is used to illuminate the sample chamber. The illumination light is separated from the trapping laser using a short-pass filter before it enters the camera. Two feedback loops continuously regulate the independent AODs to compensate for any fluctuation in the total intensity read by the photodiodes (PDs).  A LabVIEW-based FPGA data acquisition system (NI 7855R) collects the voltage signals from QPD and sends the command signals to AODs. The FPGA card runs the control protocol with a deterministic time step of 6 $\mu$s.

To create dynamics that are better than only qualitatively correct requires careful calibration. We use a three-step process.  First, using the camera as a length standard, we calibrate the image pixel size.  Next, we calibrate the trap displacement due to the change in the modulation voltage in the AOD.  Having determined the camera and AOD calibration constants, we calibrate the response of the QPD against the AOD modulation voltage.  A linear calibration for the QPD holds for a very small range ($\approx 100$ nm). Since the linear range of the AOD displacements ($\approx 400$ nm) is larger than the QPD linear range, the measured linear range is limited by the latter.  Finally, we calibrate the force exerted by the trap on the particle. The trapping force is linear over $\pm45$ nm from the trap center.  The maximum force applied was 1.5 pN for a beam power of 50 mW at the back aperture of the objective.  The calibration process for QPD and force has to be repeated each time before measurement. One final calibration is to offset the small trapping force ($k\approx$ 1 pN/$\mu$m) due to the detection laser.  We do so by creating a region of nominally zero potential and subtracting the residual small, mostly parabolic potential.  The force from this fixed potential is compensated for when determining the force applied via the potential offset at each time step of trap operation.

\begin{figure}[tbh]
	\includegraphics[width=3.0in]{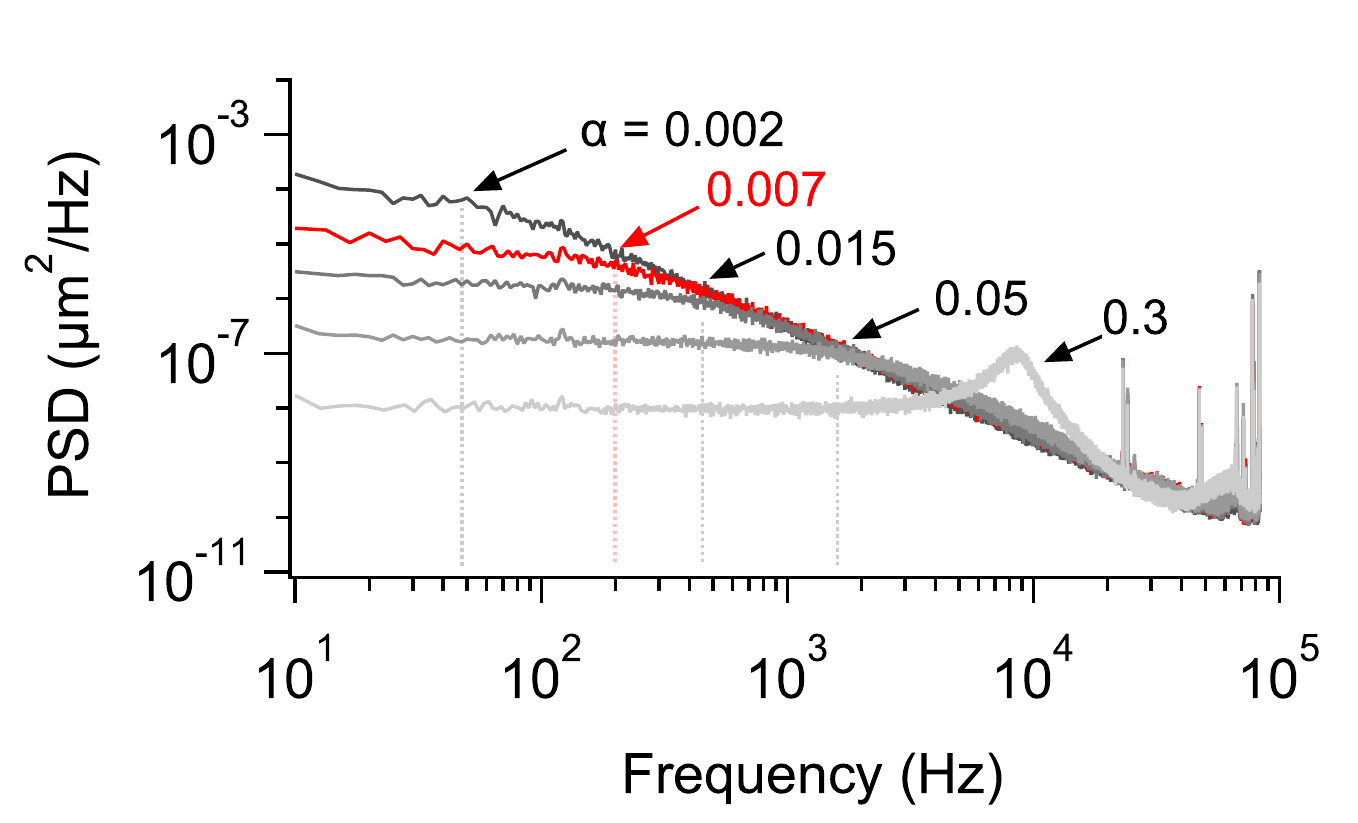}
	\caption[PS] {\label{fig:PS} 
Power spectral density for different values of proportional feedback gain ($\alpha = \Delta t/t_{\rm r}$). The power spectrum for the ``natural'' trap without feedback is shown in red. Spurious peaks in the power spectrum at frequencies $> 20$ kHz result from intensity fluctuations of the laser.}
\end{figure}

In Figure \ref{fig:PS}, we show that the feedback trap can alter the effective stiffness, $k$ of the virtual potential.  For a total delay $t_{\rm d} \approx 3 \Delta t$, the discrete dynamics of the particle in a virtual potential created by optical tweezers follows
\begin{subequations}
\begin{align}
	x_{n+1} &= x_n -\tilde{\alpha} \overline{x}_n + \xi_n \\
	\overline{x}_{n+1} &= x_{n-2} + \zeta_n \\
	x^{\rm trap}_{n+1} &= \overline{x}_{n+1}(1-G) \,,
\end{align}
\label{eqn:update rule}
\end{subequations}
\noindent where $x_n$ is the true position of the particle, $\bar{x}_n$ the observed position, $x_n^{\text{trap}}$ the trap position at time $t_n$ and $\xi_n$ and $\zeta_n$ reflect integrated thermal and measurement noise.\cite{jun2012virtual}  The gain $G = k / k_{\rm trap}$ is the ratio of stiffnesses of the desired feedback trap to the usual tweezer force constant.  Notice that $G=1$ implies placing the trap at $x=0$:  this is the usual operation of optical tweezers.

The trap constant $\tilde{\alpha} = (k/\gamma) t_{\rm trap} 
[1-\exp (-\Delta t/t_{\rm trap})]$ is dimensionless and reflects the relaxation in the tweezer during the feedback loop update time interval $\Delta t$.  Here, $\gamma \approx 6 \pi \eta r$ is the Stokes-Einstein dissipation for a particle of radius $r$ in a fluid of viscosity $\eta$.  For $\Delta t \ll t_{\rm trap}$, $\tilde{\alpha} \approx (k/ \gamma ) \Delta t \equiv \alpha$, the usual result for constant-force feedback traps.  For $\Delta t \gg t_{\rm trap}$, we have $\tilde{\alpha} \approx (k/\gamma) t_{\rm trap}$.  We operate the trap in the first limit, where $\tilde{\alpha} \approx \alpha$.

The data is sampled at $\Delta t = 6\,\mu$s and the feedback delay time, including the delay from the AOD electronics (10 $\mu$s), is $t_\text{d} = 16\,\mu$s $\approx 2.7 \Delta t$.  The equations for non-integer delays are slightly more complicated.\cite{jun2012virtual} 


For larger values of $\alpha$, the particle starts to oscillate because of overcorrection of the perturbations, as indicated by the emergence of the peak in the power spectrum.\cite{jun12}  The motion is undesirable both for the longer relaxation time created by the oscillations and for the greater variance in the particle position.  The frequency at which the resonance appears depends on the time delay ($t_\text{d})$ of the feedback loop.   In Fig.~\ref{fig:PS}, we exclude the resonant $\alpha=0.3$ curve from our estimate of bandwidth increase.

We create a static virtual double-well with our feedback trap. Such potentials have previously been created with a rapidly scanning single-beam optical tweezers between two positions.\cite{visscher1993micromanipulation}  However, multiplexed optical tweezers can impose only a limited range of potentials.  Here, we define a double-well potential from three parabolic pieces that are joined together in a way that makes the function and its first derivative continuous but has two jump discontinuities in the second derivative. The parametric form allows independent control of well separation and barrier height.  To simplify the equations, we scale energy by $k_\text{B}T$ and lengths by $\sqrt{D\Delta t}$, where $D$ is the diffusion constant of the particle, and $\Delta t$ the sampling time.  Specifically, we define
 \begin{equation}
    U(x^\prime) \equiv 
    \begin{cases}
    	\tfrac{1}{2}\alpha(x^\prime+x_{\rm m}^\prime)^2  &x^\prime \leq -x_{\rm p}^\prime \\[3pt]
	\tfrac{1}{2}\left[\tfrac{2E_{\rm b}}{\left({x_{\rm m}^\prime}^2 
		-\tfrac{2E_{\rm b}}{\alpha}\right)}\right]x^2  
		&-x_{\rm p}^\prime < x^\prime < x_{\rm p}^\prime \\[6pt]
    \tfrac{1}{2}\alpha(x^\prime-x_{\rm m}^\prime)^2  &x^\prime \geq -x_{\rm p}^\prime
    \end{cases}
\label{eqn:pieceDW}
\end{equation}

\noindent where $x_{\rm m}^\prime$ is the well position, $\alpha = 0.03$ the proportional feedback constant near the minimum of the potential well, and $E_{\rm b}$ the potential barrier.  The matching point $x_{\rm p}^\prime = \bigl( \tfrac{k_1}{k_1 + k_2} \bigr) x_{\rm m}^\prime$ and is defined by enforcing continuity of $U$ and $\partial_{x^\prime} U$.  The force constants of the stabilizing potential, $k_1$ and the destabilizing potential, $k_2$ can be calculated from the values of $\alpha$ and $E_{\rm b}$.  This parametrization of a double-well potential is more flexible than the one used in Ref.\citenum{kumar2018}, as we can independently control the well separation and barrier height, keeping the curvature of the wells fixed.  

\begin{figure}[h!]
	\includegraphics[width=3.0in]{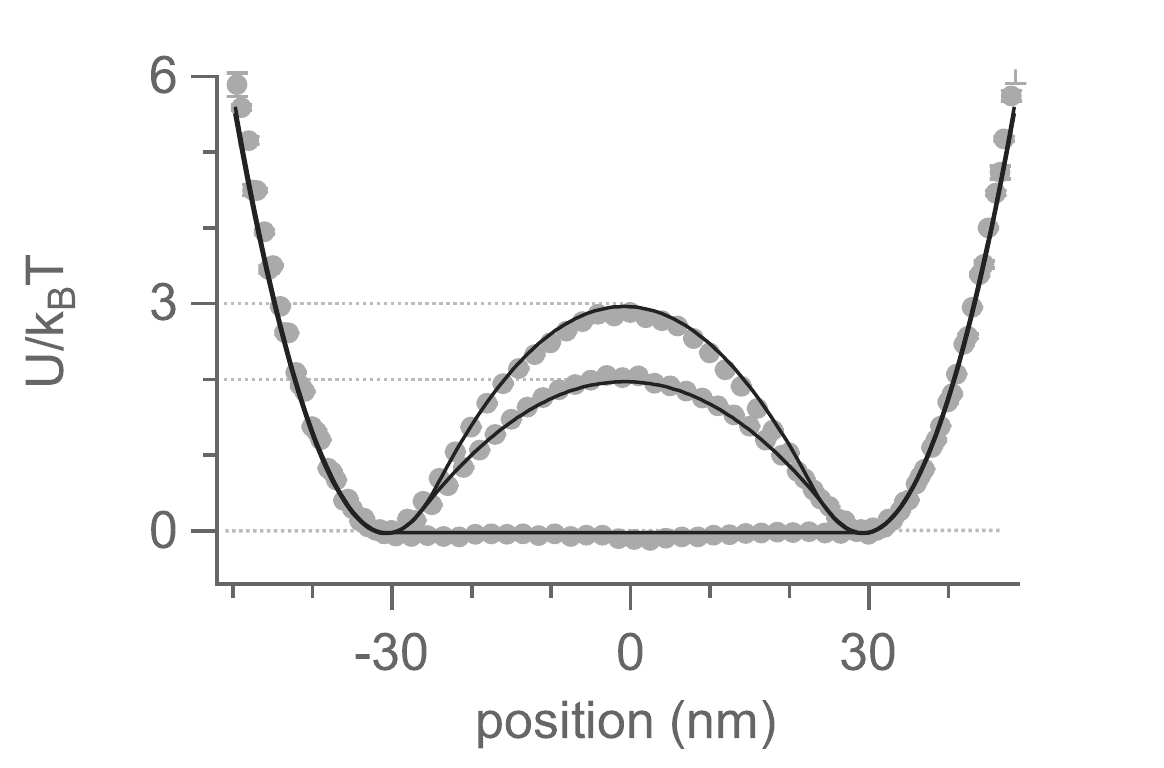}
	\caption[dw] { \label{fig:dw} 
Virtual double-well potentials with different barrier heights ($E_\text{b}/{k_{\text{B}}T} =$ 0, 2, and 3) and fixed well separation.  Gray markers denote potentials reconstructed from the Boltzmann distribution of the position measurements, with no curve fitting; the  superimposed solid black lines show the imposed potentials.  Sampling time $\Delta t = 6~\mu$s.  Time series duration is 50 s for all three cases.}
\end{figure}

Figure~\ref{fig:dw} shows a family of double-well potential curves reconstructed from their respective time series using the Boltzmann distribution, $p(x^\prime) \sim \exp [-U(x^\prime)/k_\text{B}T ]$, where the well separation is 60 nm.  Although the curves in Fig.~\ref{fig:dw} are plotted, not fit, we have confirmed that best-fit values for parameters such as $E_\text{b}$ and $x_\text{m}$ are within 5$\%$ of the values imposed by the control program.

\begin{figure}[b!]
	\includegraphics[width=3.0in]{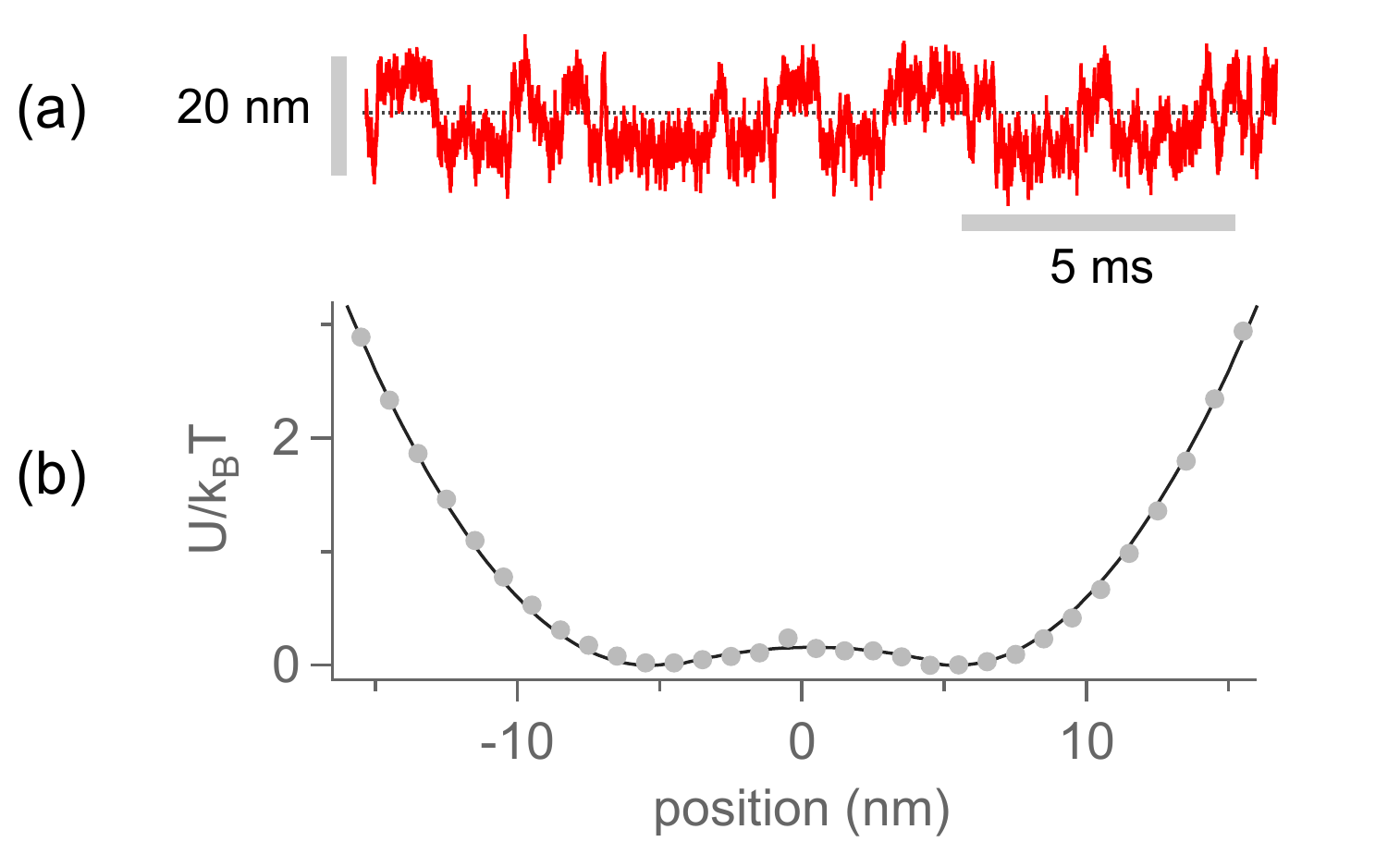}
	\caption[dw_min] { \label{fig:dw_min} 
(a) Time series showing spontaneous hops between two states. (b) The reconstructed potential (gray markers) shows wells 10.6 nm apart and a barrier height 0.16 $k_{\text{B}}T$. Black solid line is a fit using Eq.~\ref{eqn:pieceDW}.  }
\end{figure}

Another important feature of a feedback trap is that the scale of potentials is not limited by the optical resolution of the microscope.  In Figure~\ref{fig:dw}, the well separation was 60 nm.  To see how small a scale we could create such a potential, we abandoned our limitation of feedback gain\cite{jun2012virtual} to $\alpha < 0.038$, which is necessary for the feedback trap to imitate closely both the statics and dynamics of the imposed potential.  We chose $\alpha \approx 0.014$, which roughly corresponds to critical damping.

Figure \ref{fig:dw_min}a shows the time series for smallest well separation that we could achieve with feedback under these conditions. The well separation is $10.6$ nm, which is far below the diffraction limit $\approx 220$ nm (Fig. \ref{fig:dw_min}b).  Time-shared traps cannot create potentials with independently tunable barrier height, well separation, and well curvature.  Spatial light modulators cannot create these shapes at sub-diffractive-limit length scales.  Although the energy barrier is quite low (0.16 $k_{\rm B}T$), the small curvature of the barrier still leads to two-state behavior in the time series, where the dwell time in a well is $\approx 10$X the transition time between wells.

In an optical trap, the axial stiffness is smaller than the transverse stiffness because of radiation pressure and the weaker gradient of intensity along the axis of the focused laser beam.\cite{jones15}  Anisotropic traps used as force sensors have the disadvantage that the measurement bandwidth differs according to the direction of force that is applied.  An isotropic trap would allow unbiased measurement of dynamics in a three-dimensional environment.

\begin{figure}[h]
	\includegraphics[width=3.2in]{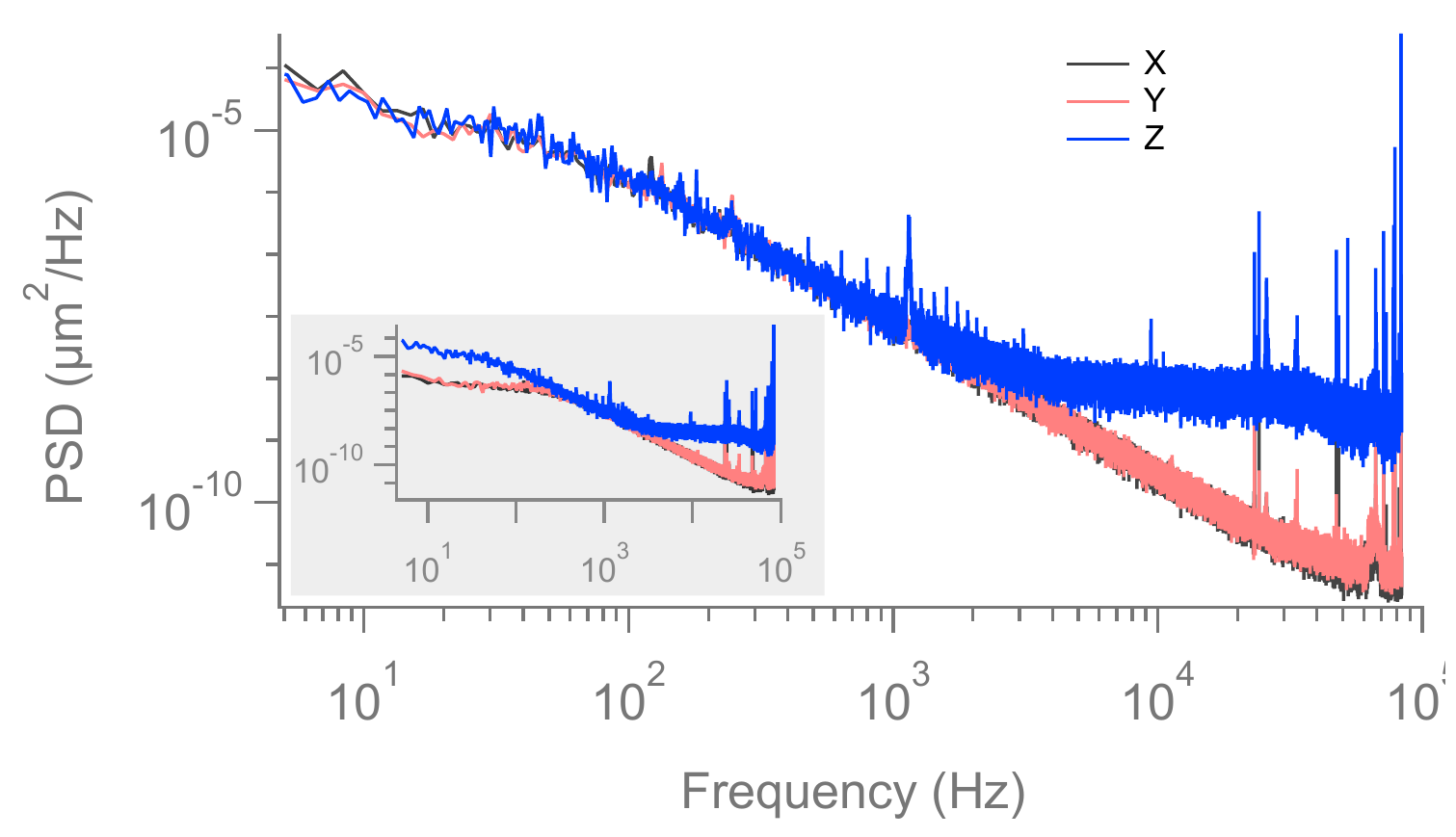}
	\caption[XYZ] { \label{fig:XYZ} 
Power spectrum density of the $x$, $y$, and $z$ signals after the feedback. (Inset: Power spectrum density of the $x$, $y$ and $z$ signals before the feedback.)}
\end{figure}

Here, we show that we can use feedback to \textit{reduce} the lateral stiffness of the trap to make the trap isotropic, with equal stiffness in the lateral and axial directions   (Fig.~\ref{fig:XYZ}). The axial position of the trapped particle is estimated from the fluctuation in total intensity on the detector.  Currently, the axial sensitivity is limited to a smaller bandwidth ($\approx 2$ kHz) as compared to transverse sensitivity ($\approx 60$ kHz).  With improved axial sensitivity, it should be possible to create an isotropic trap by \textit{increasing} the axial stiffness to match the lateral stiffness.

In this letter, we have demonstrated a feedback trap system based on optical tweezers.  Using this feedback technique, we have created virtual harmonic potentials with different stiffness constants, static double wells with independently tunable parameters, and isotropic traps.  These applications of optical tweezers have not been shown before using techniques such as time-shared traps.  Although our work is for two-dimensional traps, it can be readily extended to feedback in three dimensions, using the intensity to measure axial position (as done here) or using a variety of more sophisticated techniques\cite{dreyer2004improved,deufel2006detection,shechtman14} and a method to move the trap position axially. 

The ability to create and control energy landscapes at scales comparable to the size of proteins offers intriguing possibilities for biophysical applications.  For example, recent experiments suggest that some protein folding is well described by diffusive dynamics on an effectively one-dimensional energy surface.\cite{neupane16}  Using the techniques developed here, one could create model systems with similar dynamics.  Also, one could place a colloidal particle in a potential whose dynamics could imitate, in a controllable way, the dynamics of a ligand.  Even more intriguingly, those dynamics could be adaptive, allowing exploration of phenomena such as catch bonds, whose dissociation lifetime increases sharply when pulled.\cite{thomas02}  Such studies would likely be facilitated by using smaller particles.  Techniques such as interferometric scattering microscopy (iSCAT) have shown that by interfering a reference beam with scattered light, it is possible to detect colloidal particles and even proteins on a 10-nm scale.\cite{young18}  Reducing the delays and feedback latency will allow a further reduction in the scale of potentials.

Finally, as discussed in the introduction, time-dependent potentials can be used to carry out interesting stochastic thermodynamic experiments.  With feedback bandwidths 1000X faster than our previous work on slow stochastic processes,\cite{jun2014high,gavrilov16b,gavrilov17b} we can address problems with faster dynamics such as finite-time transformations in non-equilibrium thermodynamics.\cite{aurell2012refined,zulkowski2014optimal}  Such experiments could also take advantage of another feature of feedback traps based on optical tweezers:  because the applied forces are localized (in contrast to traps based on electrokinetic forces), they allow custom energy landscapes containing multiple particles. 

We are grateful to David Sivak and Nancy Forde for comments. This work was funded by Discovery and RTI Grants from NSERC (Canada).

\bibliographystyle{apsrev4-1}
\bibliography{APL_ref}

\end{document}